\documentclass[aps,prb,twocolumn]{revtex4-1}
\usepackage{bm}
\usepackage{graphicx}
\usepackage{graphics}
\usepackage{amsmath,amssymb,amstext}
\usepackage{amsfonts}

\usepackage{epstopdf}
\usepackage{hyperref}
\hypersetup{
    colorlinks,%
    citecolor=blue,%
    linkcolor=blue,%
    urlcolor=blue
}
\usepackage{tikz}\usetikzlibrary{petri}

\begin{document}

\newcommand{\be}   {\begin{equation}}
\newcommand{\ee}   {\end{equation}}
\newcommand{\ba}   {\begin{eqnarray}}
\newcommand{\ea}   {\end{eqnarray}}
\newcommand{\ve}  {\varepsilon}

\newcommand{\nhat}{\hat{n}}
\newcommand{\veck}{\textbf{k}}
\newcommand\ep{\epsilon}
\newcommand\g{\gamma}
\newcommand\s{\sigma}
\newcommand\up{\uparrow}
\newcommand\dw{\downarrow}
\newcommand\down{\downarrow}
\newcommand{\ed}[1]{\ep_{d#1}}
\newcommand{\ket}[1]{\vert #1 \rangle}
\newcommand{\ann}{a^{\dagger}}
\newcommand{\dann}{d^{\dagger}}
\newcommand{\tdots}{t_{dots}}
\newcommand{\gammaA}[1]{\gamma_{A,#1}}
\newcommand{\gammaB}[1]{\gamma_{B,#1}}
\newcommand{\Green}[1]{G_{#1}(\omega) }

\newcommand{\GreenG}[2]{G_{#1}^{ #2} (\omega) }

\newcommand{\super}{\vert \Delta \vert}

\title{ Manipulating Majorana zero modes in double quantum dots }

\author{Jesus D. Cifuentes}
\affiliation{Instituto de F\'{\i}sica, Universidade de S\~{a}o Paulo,
C.P.\ 66318, 05315--970 S\~{a}o Paulo, SP, Brazil}
\author{Luis G.~G.~V. Dias da Silva}
\affiliation{Instituto de F\'{\i}sica, Universidade de S\~{a}o Paulo,
C.P.\ 66318, 05315--970 S\~{a}o Paulo, SP, Brazil}

\date{ \today }

\begin{abstract}
Majorana zero modes (MZMs) emerging at the edges of topological superconducting wires have been proposed as the building blocks of novel, fault-tolerant quantum computation protocols. Coherent detection and manipulation of such states in scalable devices are, therefore, essential in these applications. Recent detection proposals include semiconductor quantum dots (QDs) coupled to the end of these wires, as changes in the QD electronic spectral density due to the MZM coupling could be detected in transport experiments. Here, we propose that multi-QD systems can also be used to \textit{manipulate} MZMs through precise control over the QDs' parameters. The simplest case where Majorana manipulation is possible is in a double quantum dot (DQD) geometry.  By using exact analytical methods and numerical renormalization-group calculations, we show that the QDs' spectral functions can be used to characterize the presence or not of MZMs ``leaking'' into the DQD. More importantly, we find that these signatures respond to changes in the DQD parameters such as gate-voltages and couplings in a consistent fashion. Additionally, we show that different MZM-DQD coupling geometries (``symmetric'' , ``in-series'' and ``T-shaped'' junctions) offer distinct ways in which MZMs can be switched from dot to dot. These results highlight the interesting possibilities that DQDs offer for all-electrical MZM control in scalable devices.

\end{abstract} 

\maketitle



\maketitle

\section{Introduction}
\label{sec:Intro}

The search for Majorana quasiparticles in condensed matter systems has gained renewed attention in the last decade, motivated  by the exciting prospects for achieving scalable, fault-tolerant topological quantum computation protocols in nanoscale devices \cite{nayak_non-abelian_2008,alicea_new_2012,beenakker_search_2013,Aguado::40:523:2017}. From the landmark theoretical proposals for realizing Kitaev's model of a 1D topological superconductor \cite{kitaev_unpaired_2001,kitaev_fault-tolerant_2003} using semiconductor quantum wires with strong-spin orbit coupling and proximity-induced superconductivity  \cite{lutchyn_majorana_2010, oreg_helical_2010}, the field  rapidly evolved towards the first experiments reporting data consistent with these predictions \citep{mourik_signatures_2012,das_zero-bias_2012,deng_anomalous_2012}. As a result, the last few years have been full of excitement  as continuing improvements in sample growth and characterization techniques  have allowed for more consistent experimental evidence for Majorana bound states in semiconductor quantum wires \cite{deng_majorana_2016,zhang_quantized_2018,Deng:Phys.Rev.B:98:085125:2018}.

In these set-ups, ``Majorana signatures''  are characterized by zero-bias signals in the conductance across the device due to the emergence of robust zero-energy modes localized at the edges of the quantum wire. An important requirement is to distinguish these so-called Majorana zero modes (MZM) from other zero-bias phenomena, such as the Kondo effect \cite{hewson_kondo_1997}, which have been found in similar systems \cite{lee_zero-bias_2012}. A large effort in recent experimental proposals was put on ways to uniquely identify MZMs, including proposals for measuring the signatures of non-Abelian statistics \cite{aasen_milestones_2016,sarma_majorana_2015,heck_coulomb-assisted_2012}. Although this last property is crucial in the implementation of fault-tolerant quantum computers, its measurement has been elusive so far as it requires overcoming several experimental issues related to the need of ``moving'' Majorana quasiparticles in order to perform braiding operations \cite{Hoffman:Phys.Rev.B:045316:2016,karzig_scalable_2017}.

A rather straightforward proposal to detect MZMs consists of attaching a quantum dot (QD) to the edge of a topological quantum wire and then measuring the electrical conductance through the QD \cite{liu_detecting_2011}.  In such an arrangement, the MZM at the end of the chain ``leaks'' into the attached  QD \cite{vernek_subtle_2014}, reducing the zero-bias conductance through the dot by the sizable amount of $\frac{e^{2}}{2h}$.  This detection method offers two key advantages over other approaches: (i) no direct charge transfer between the MBS and the dot is necessary, thus preventing ``quasiparticle poisoning'' \cite{Rainis:Phys.Rev.B:85:174533:2012}, and (ii) a clear distinction with Kondo physics is warranted, even if the experiment is performed at temperatures below the  Kondo temperature $T_K$ \cite{lee_kondo_2013,ruiz-tijerina_interaction_2015,gorski_interplay_2018}. Recently, topological quantum wire-QD junctions  have been realized in experiments \cite{deng_majorana_2016,Deng:Phys.Rev.B:98:085125:2018}, paving the way for further experiments involving the detection of MZMs using quantum dots.

Venturing beyond simple ``detection'' setups, the large degree of control over the QD parameters offers the unique possibility of \textit{manipulating} MZMs inside multidot systems. The simplest case where Majorana manipulation is possible is in a double quantum dot (DQD). Tunneling Majorana modes in these basic structures have inspired theoretical studies \cite{silva_andreev_2016,ivanov_coherent_2017,Loss2019,Ricco:Phys.Rev.B:99:155159:2019}  and experimental setups confirming the observations of Andreev molecules \cite{su_andreev_2017}. 
However, despite the fact that DQDs offer several possibilities for manipulation of MZMs, there is still no complete analysis of the possible transitions of these Majorana signatures between the QDs even in a simple model.

 In this paper, we explore the different possibilities for Majorana manipulation in a device consisting of a DQD coupled to a MZM and a metallic lead (see Fig.\ \ref{fig:GenModel}). The simplicity of this model allows us to analytically explore different geometries of QD's from symmetric and ``in-series'' couplings to T-shaped junctions (Fig.\ \ref{fig:MajoranaModels}). We considered both noninteracting and interacting regimes, observing major agreement between both approaches about the location of the Majorana signature.
 
 We performed a detailed study of the noninteracting DQD limit, by using Zubarev's procedure  \cite{zubarev_double-time_1960} to provide an exact formula to calculate the spectral functions. For the interacting case, we resort to numerical renormalization group (NRG)\cite{bulla_numerical_2008} calculations for this model.
 While the noninteracting regime is suitable to obtain exact expressions for the Green function, the interacting case  shows how the Majorana signature co-exists with strongly correlated phenomena such as the Kondo effect \cite{hewson_kondo_1997} and RKKY interactions  \cite{ruderman_indirect_1954,kasuya_theory_1956,yosida_magnetic_1957}.

This paper is organized as follows. In Sec.\ \ref{sec:modelmethods} we describe the model of a DQD coupled to a MZM and to a metallic lead, as well as the methods used.  The results are presented in Sec.\ \ref{sec:results} where we compare the noninteracting density of states (LDOS) (Sec.\ \ref{subsect:non-int}) with the low-energy  interacting results in Sec.\ \ref{subsec:Interacting}. Finally, our conclusions are given in Sec.\ \ref{sec:Conclusions}.

\section{Model and methods}
\label{sec:modelmethods}

\begin{figure}[t]
\begin{center}
\includegraphics[scale=0.4]{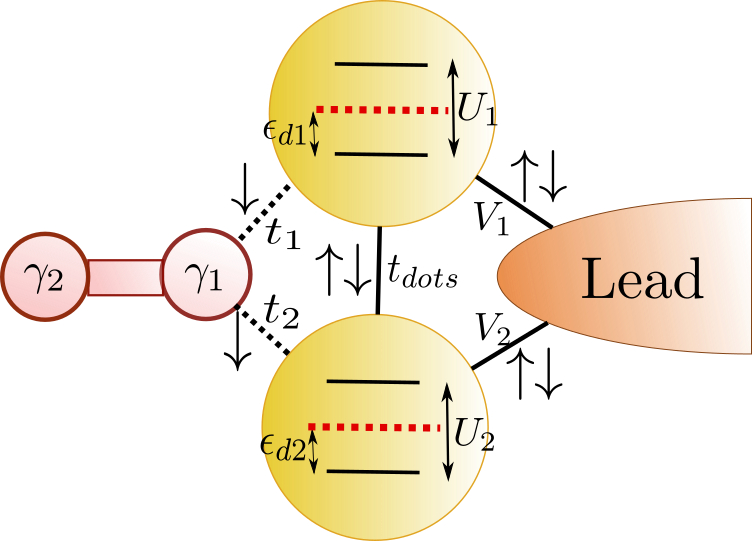}
\end{center}
\caption{ Model for the DQD-Majorana system. Solid lines represent the hoppings ($t_{dots}$: interdot coupling , $V_1,V_2$ couplings of QD1 and QD2 with the lead. ). Dashed lines: MZM-DQD spin-down effective couplings $t_1,t_2$. The atomic energy levels appear inside each QD $\ep_1, \ep_2$ are tuned by the gate voltages. The Coulomb interaction in each dot is represented by $U_1,U_2$.  The red dashed horizontal lines represent the Fermi level of the metallic lead.
}
\label{fig:GenModel}
\end{figure}

We consider the setup shown in Fig.\ \ref{fig:GenModel}, in which a single MZM $\gamma_1$ located at the edge of a 1D topological superconductor is coupled to a double quantum dot (DQD) attached to a single metallic lead. The Hamiltonian of the entire system can be expressed as:
\begin{equation}
H=H_{\rm DQD}+H_{\rm lead}+H_{\rm DQD-lead}+H_{\rm M-DQD} 
\label{eq:Model}
\end{equation}
where the different terms describe, respectively, the (interacting) DQD, the (noninteracting) metallic lead, and the DQD-lead couplings, written as:
\begin{align}
    H_{\rm DQD} =& \!  \sum_{\substack{i\!=\! 1,2 \\ \sigma \!=\! \dw , \up }} \left(\epsilon_{i}+\frac{U_i}{2}\right)\hat{n}_{i\sigma} + \frac{U_i}{2}\left(\sum_{\sigma} \hat{n}_{i\sigma}-1\right)^{2} \nonumber \\ 
& + \sum_{\sigma} \tdots(\dann_{1\sigma}  d_{2\sigma}+\dann_{2\sigma}  d_{1\sigma}) \; ,   \nonumber  \\ 
H_{\rm lead} =&  \sum_{\mathbf{k}\sigma }\epsilon_{\mathbf{k}}c_{\mathbf{k}\sigma}^{\dagger}c_{\mathbf{k}\sigma }  \; ,   \nonumber \\ 
H_{\rm DQD-lead} =&  \sum_{\mathbf{k}\sigma }\sum_{i\!=\!1,2 }V_{i\textbf{k}} c_{\mathbf{k}\sigma }^{\dagger}d_{i\sigma}+V^*_{i\textbf{k}} d_{i\sigma}^{\dagger}c_{\mathbf{k}\sigma } \; ,
\label{eq:H_DQD} 
\end{align}
while the DQD-MZM coupling is given by \cite{liu_detecting_2011,lee_kondo_2013,ruiz-tijerina_interaction_2015,Hoffman:Phys.Rev.B:045440:2017,Prada:Phys.Rev.B:96:085418:2017}:
\begin{equation}
H_{\rm M-DQD} =  \sum_{\sigma, i=1}^2 t_{i \sigma} \left(d_{i\sigma }^{\dagger}\gamma_{1}+\gamma_{1}d_{i\sigma}\right) \; .
\label{eq:H_MDQD_1} 
\end{equation}

In the equations above, $\epsilon_{i}$ is the energy level of dot $i$, $U_i$ is the Coulomb repulsion and $\tdots$ is the coupling parameter between both QDs. The operator $\dann_{i\sigma}$ creates a particle in dot $i$ with spin $\sigma$ and $\hat{n}_{i\sigma}=d_{i\sigma}^{\dagger}d_{i\sigma}$ is the particle number operator of state $i$,   $c_{\mathbf{k}\sigma }^{\dagger}$ is the creation operator a particle with momentum $\mathbf{k}$ and spin
$\sigma$ in the lead.  Finally, $\epsilon_{\mathbf{k}l}$ is the corresponding energy
 and $V_i(\textbf{k})$ describes the tunneling coupling between the lead and dot $i$. 

We take the length of the wire to be large so that we can safely neglect both the overalap between the two Majorana modes and the (much smaller) coupling between the DQD and the $\gamma_2$ MZM located at the other edge of the wire. We also note that the DQD-$\gamma_1$ coupling strength $t_{\sigma i}$ in Eq.\ (\ref{eq:H_MDQD_1}) above is, in general, spin-dependent \cite{Hoffman:Phys.Rev.B:045440:2017,Prada:Phys.Rev.B:96:085418:2017,Deng:Phys.Rev.B:98:085125:2018} and can be written  in terms of the $\gamma_1$ MZM ``spin canting angle'' $\theta_1$  as  $\left(t_{\up i},t_{\dw i} \right) \equiv t_{i} \left(\sin{\frac{\theta_1}{2}}, -\cos{\frac{\theta_1}{2}}\right)$. For the purposes of this work, we take  $\theta_1 \! = \!\pi$  such that only spin down dot operators are coupled to the MZM, making $H_{\rm M-DQD}$ fully spin-conserving. This choice adds an extra symmetry (spin down parity) to the full Hamiltonian, which will turn out to be important in the NRG calculations presented in Sec.\ \ref{subsec:Interacting}.  

It is also useful to recast the last term of Eq.\ (\ref{eq:Model}) in terms of (Dirac) fermionic operators. Following Refs. \onlinecite{lee_kondo_2013,ruiz-tijerina_interaction_2015}, we choose to write the MZMs $\gamma_1$ and $\gamma_2$ as a superposition of the creation ($f^{\dagger}_\dw$) and annihilation ($f^{ }_\dw$) operators of a (fully polarized) spin down fermion:
\begin{equation}
    \gamma_1 = \frac{1}{\sqrt{2}} \left( f^\dagger_{\dw} + f_{\dw}\ \right) \; , \gamma_2 = \frac{i}{\sqrt{2}} \left( f^\dagger_{\dw} - f_{\dw} \right) \; . \label{eq:MajOp}
\end{equation}

In this representation, the effective coupling between the MZM and the DQD given by Eq.\ (\ref{eq:H_MDQD_1}) becomes:
\begin{eqnarray}
    H_{\rm M-DQD} & = &  
\sum_{i}t_{i} \left(d_{i\downarrow}^{\dagger}f^\dagger_{\dw} + 
     f_{\downarrow}d_{i\dw} +d_{i\downarrow}^{\dagger}f_{\dw}+
     f_{\downarrow}^{\dagger} d_{i\downarrow}\right) \; , 
    \label{eq:H_MDQD}
\end{eqnarray}
where $t_i \equiv t_{\dw i}$ is the coupling strength between the MZM and QD $i$.

In order to identify the presence/absence of MZMs ``leaking'' from the edge of the TS into the dots \cite{liu_detecting_2011,vernek_subtle_2014,lee_kondo_2013,ruiz-tijerina_interaction_2015}, the quantities of interest are the spin-resolved spectral functions (or, equivalently, the local density of states) of the quantum dots. As usual, the spectral function for spin $\sigma$ in dot $i$ is defined as:
\begin{equation}
    \rho_{i \sigma}(\omega)\equiv-\frac{1}{\pi} \textrm{Im} \left[G_{d_{i \sigma},d_{i \sigma}^\dagger}(\omega)\right].
    \label{eq:SpecFunc}
\end{equation}
where $G_{d_{i \sigma},d_{i \sigma}^\dagger}(\omega) \equiv \langle\langle d_{i \sigma},d_{i \sigma}^{\dagger} \rangle \rangle_\omega$ is the retarded (diagonal) Green's function involving dot $i$ operators $d_{i \sigma}$ and $d_{i \sigma}^\dagger$. Next, we describe the procedures for calculating  $\rho_{i \sigma}(\omega)$ in the regimes of weak ($U_i \ll V, t_i$) and strong ($U_i \! \gg\! V, t_i$) electron-electron interaction in the dots.


\subsection{Noninteracting limit: Graph-Gauss-Jordan elimination }
\label{sec:noninteractingMethods}

In the noninteracting limit ($U_i\!=\!0$), the Hamiltonian $H$ is quadratic in the fermionic operators and we can obtain analytic expressions for the spectral densities defined in Eq.\ (\ref{eq:SpecFunc}). Using Zubarev's equation of motion (EOM) approach \cite{zubarev_double-time_1960}, we can derive  exact expressions for the Green functions associated to both quantum dot operators $(\Green{d_1d^\dagger_1},\Green{d_2d^\dagger_2})$. 

 The EOM  equations define a $8 \times 8$ linear system where the Hamiltonian parameters $(t_1,t_2,\epsilon_1 \ldots)$ and the energy $\omega$ are taken as algebraic variables. The solution for these types of equations is a finite continued fraction of multivariate polynomials with maximum degree $8$, which makes it difficult to provide an exact solution using either analytic or numerical methods. To bypass this problem, we introduce a Graph-Gauss-Jordan elimination process \cite{spielman_algorithms_2010} to iteratively solve the coupled equations of motion. We briefly describe the procedure here. 

We begin by representing the Majorana-DQD quantum dot system in a ``flow graph'', where each spin-resolved fermion operator (e.g., $d^{\dagger}_{1 \dw}$, $d_{1 \dw}$, $f_{\dw}$, $f^{\dagger}_{\dw}$, etc.) is represented as a ``vertex". The coefficients of the quadratic terms (such as $d^{\dagger}_{1 \dw} d_{1 \dw}$ or $c^{\dagger}_{k \dw} c_{k\dw}$, etc.) are associated to each node as ``self-energies'' while the coupling terms involving two fermion operators  (such as $d^{\dagger}_{1 \dw} f_{\dw}$ or $d^{\dagger}_{1 \dw} f^{\dagger}_{\dw}$, etc.) are associated to the ``links''  connecting the respective vertices (see Fig.\ \ref{fig:GaussJordanGraph} in the Appendix). 

We then proceed to iteratively remove both vertices and links by rewriting the self-energies and couplings in terms of the eliminated variables, such that each vertex elimination depicts another step in the  Gauss-Jordan process. In the end, the self-energy of the only remaining vertex will contain the full information needed in order to compute the target Green's function.

This method proved to be efficient in solving complex systems of coupled Green's functions as the graph elimination process provides a natural linear algorithm to compute the targeted continued fraction. Moreover, the graphic representation simplifies the procedure and allows one to readily identify minimal coupling points, which could reduce the complexity of the solution. A detailed description of the method is given in Appendix \ref{sec:Appendix_alg}.

After applying the Graph-Gauss-Jordan process, we obtain a closed form for the noninteracting Green's functions. For instance, the GF for dot 1 (which is directly coupled to the MZM) will be given by:
\begin{equation}
G_{{d_{1\downarrow},d_{1\downarrow}^{\dagger}}}\left(\omega\right)=\frac{1}{\omega-\epsilon_{DQD}^{+}-\frac{\left\Vert T_{+}\right\Vert ^{2}}{\omega-\epsilon_{M}-\frac{\left\Vert T_{-}\right\Vert ^{2}}{\omega -\epsilon_{DQD}^{-}}}} \; ,
    \label{eq:Green_NonInteracting}
\end{equation}
where the energies $\epsilon_{DQD}^{\pm}$ are given by
\begin{equation}
\epsilon_{DQD}^{\pm}=\pm\epsilon_{1}+\sum_{\mathbf{k}}\frac{V_{1}V_{1}^{*}}{\omega-\epsilon_{\mathbf{k}}}+\frac{\left\Vert \pm t_{dots}+\sum_{\mathbf{k}}\frac{V_{1}V_{2}^{*}}{\omega-\epsilon_{\mathbf{k}}}\right\Vert ^{2}}{\omega\mp\epsilon_{2}-\sum_{\mathbf{k}}\frac{V_{2}V_{2}^{*}}{\omega-\epsilon_{\mathbf{k}}}} \; , \label{eq:epDQD}
\end{equation}
\noindent 

\begin{equation}
    T_{\pm}=\pm t_{1}\pm t_{2}\frac{\left(\pm t_{dots}+\sum_{\mathbf{k}}\frac{V_{1}V_{2}^{*}}{\omega-\epsilon_{\mathbf{k}}}\right)}{\omega\mp\epsilon_{2}-\sum_{\mathbf{k}}\frac{V_{2}V_{2}^{*}}{\omega-\epsilon_{\mathbf{k}}}} \; , \label{eq:T+-}
\end{equation}
and
\begin{equation}
    \epsilon_{M}=\omega-\frac{\left\Vert t_{2}\right\Vert ^{2} } {\omega-\epsilon_{2}-\sum_{\mathbf{k}}\frac{V_{2}V_{2}^{*}}{\omega-\epsilon_{\mathbf{k}}}}-\frac{\left\Vert t_{2}\right\Vert ^{2}}{\omega+\epsilon_{2}-\sum_{\mathbf{k}}\frac{V_{2}V_{2}^{*}}{\omega+\epsilon_{\mathbf{k}}}} \; . \label{eq:M2}
\end{equation}

The spin-up spectral density, which is \textit{not} coupled to the MZM, can be obtained by taking $t_1,t_2 = 0$ in Eqs.\ (\ref{eq:Green_NonInteracting})-(\ref{eq:M2}), hence giving
\begin{equation}
    G_{{d_{1\uparrow},d_{1\uparrow}^{\dagger}}}\left(\omega\right)=\frac{1}{\omega-\epsilon_{DQD}^{+}} \; .
    \label{eq:Green_NonInteractingII}
\end{equation}

The final results will depend on the broadening parameter of QD $i$ with the lead $(\Gamma_i)$, given, in the broad-band limit, by:

\begin{equation}
   -i\Gamma_i = \lim_{s\rightarrow 0} \sum_{\boldsymbol{k}}\frac{V_{i}^{*}V_{i}}{\omega+ is -\epsilon_{\boldsymbol{k}}}.
\end{equation}

Finally, we compute the spin-resolved LDOS in dot 1 as:
\begin{equation}
    \rho_{1\sigma}(\omega)=-\frac{1}{\pi} \textrm{Im} \left[G_{d_{1\sigma},d_{1\sigma}^\dagger}(\omega)\right].
    \label{eq:Density of States}
\end{equation}

Similar results can be obtain for the LDOS of the second $\rho_{2\sigma}$ by exchanging the indexes $1$ and $2$ in Eq.\ \eqref{eq:Green_NonInteractingII}.

\subsection{Interacting limit: Wilson's NRG }
\label{sec:NRG-interacting}


To address the case of \textit{interacting} quantum dots, we employ the numerical renormalization group (NRG), one of the most successful methods to study interacting quantum impurity models (QIMs)  \cite{wilson_renormalization_1975,sindel_numerical_2005,bulla_numerical_2008}. In general, a QIM describes a system spanning a finite and relatively small Hilbert space (the ``impurity") coupled to a much larger system (a ``continuum"), spanning a large (typically infinite) Hilbert space. As it turns out, the Hamiltonian in Eq.\ (\ref{eq:Model}) can be cast as a QIM where the impurity is the DQD coupled to the Majorana mode, which is then coupled to the continuum of electrons in the metallic leads.

We notice that the DQD-Majorana tunneling term given by Eq.\ (\ref{eq:H_MDQD}) effectively breaks total spin $S_z$ and charge $Q$ conservation of the whole system, while it preserves spin-$\downarrow$ parity $P_{\downarrow}\!=\! \pm 1$ and spin up particle number $N_\up$. To  improve the efficiency of the method, we used these symmetries to maintain a block structure during NRG's iterative diagonalization process  \cite{bulla_numerical_2008,lee_kondo_2013,ruiz-tijerina_interaction_2015}. 
Both the states serving as a basis for the initial impurity Hamiltonian and the single-site Wilson chain states can be grouped in $(N_\up,P_\dw)$ blocks. Thus, the $(N_\up,P_\dw)$ block structure is preserved during the entire NRG iteration process \cite{bulla_numerical_2008}. In order to compute the (interacting) spectral functions, we use the density matrix NRG (DM-NRG) procedure  \cite{hofstetter_generalized_2000} in combination with the z-trick method \cite{oliveira_generalized_1994}, which improves spectral resolution at high energies. We have checked the accuracy of the results by comparing the results with the complete Fock space method \cite{Peters:Phys.Rev.B:245114:2006} for some of the parameters used.

 \section{Results \label{sec:results}}

\begin{figure}[t]
    \begin{center}
    \includegraphics[scale=0.5]{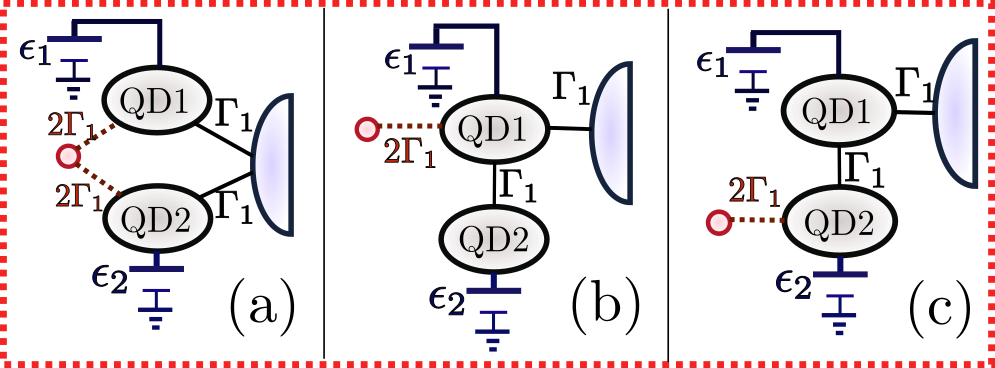}
    \caption{\label{fig:MajoranaModels}  MZM-DQD-lead coupling geometries considered. (a) Symmetric ``parallel'' MZM-DQD-lead coupling (with no interdot coupling). (b) ``T-dot''  arrangement, where dot 2 is coupled only to dot 1, and (c) MZM and quantum dots coupled ``in series'' with the lead. 
    }
\end{center}
\end{figure}

For the remainder of the paper, we will focus on the Majorana-DQD coupling geometries depicted in Fig.\ \ref{fig:MajoranaModels}: a ``symmetric coupling"  arrangement [Fig.\ \ref{fig:MajoranaModels}(a)], a ``T-shaped" configuration [Fig.\ \ref{fig:MajoranaModels}(b)] and the case where the Majorana and both dots are coupled  ``in-series" [Fig.\ \ref{fig:MajoranaModels}(c)]. As we shall see, the intensity of the MZM  signature in each dot can be controlled by external gate-voltages which change the position of the dot levels $\epsilon_{1,2}$ relative to the Fermi energy in the leads.

As mentioned previously, the spin-resolved spectral density (or local density of states LDOS) of each quantum dot provides significant information about the effective tunneling (or not) of a Majorana zero mode into the dot. By comparing the spectral densities for the cases with and without DQD-Majorana couplings, we could identify two generic types of signatures of the Majorana presence in quantum dot $i$, which we define as follows:

 \begin{itemize}
         \item \textbf{Type I: }  The spin-down LDOS is half of the spin-up LDOS  at the Fermi energy $(\rho_{i \dw}(0)=\rho_{i\up}(0)/2)$. 
         \item \textbf{Type II: } The spin-up spectral density of dot $i$ shows a zero mode of height $\rho_{i \dw}(0) = \frac{0.5}{\pi  \Gamma_1}$ while no such signature appears in the spin-up spectral density. 
     \end{itemize}
     
We should point out that the identification of MZM signatures is much simpler in the case of single-quantum dots coupled to an MZM mode considered, e.g., in Refs. \onlinecite{liu_detecting_2011,vernek_subtle_2014,ruiz-tijerina_interaction_2015}. In that case, the generic MZM signature is essentially given by the "type-II" condition defined above, which amounts to the spin-down spectral function value at the Fermi energy pinned at $0.5/(\pi  \Gamma)$. If the QD energy level is tuned to the particle-hole symmetric case, the spin up spectral function pinned at $1/(\pi  \Gamma)$ such that both ``Type-I'' and ``Type-II'' conditions apply.

In the double quantum dot set-up considered here, things are more complex as quantum interference effects give rise to situations where only one of these conditions is met. As we shall see in the following sections, either one of these  two types of signatures appear over a wide range of parameters in our results. In practice, we find that the appearance of Type-I or Type-II signatures are related to the behavior of the spin-up spectral density near the Fermi energy $\rho_{i \uparrow}(\omega\!\sim\!0)$: Type I often appears when $\rho_{i\up}(0)$ displays a peak, while Type II typically emerges in situations where $\rho_{i\up}(0) \approx 0$.

Hereafter, we shall refer to ``MZM manipulation" the changes in the Majorana signatures in the dot spectral functions induced by the tuning of the dot gate voltages $( \epsilon_1 , \epsilon_2 )$ in the three different setups depicted in Fig.\ref{fig:MajoranaModels}. In each case, we consider definite values of the couplings $\Gamma_2$, $t_{dots}$, $t_1$, and $t_2$, as follows.  In the configuration shown in Fig.\ref{fig:MajoranaModels}(a), we coupled the QD symmetrically to the lead and the MZM by setting $t_1\!=\! t_2$.  Within this setup, we expect the MZM signature to ``split" due to quantum interference and identical signatures should appear in the spectral densities of both dots. We also considered setups in which only one of the dots is coupled directly the MZM or to the metallic lead. Hence, there are only two distinct coupling geometries: either both the MZM and the lead are coupled to the same dot, forming a ``T-junction" or ``side-dot" configuration ($t_{2(1)}\!=\!0$ and $\Gamma_{2(1)}\!=\!0$), as shown in Fig.\ref{fig:MajoranaModels}(b). Alternatively, the MZM can be coupled to one of the dots and the lead to the other, such that the MZM and dots are coupled in series [$t_{1(2)}\!=\!0$ and $\Gamma_{2(1)}\!=\!0$, see Fig.\ref{fig:MajoranaModels}(c)]. In the remainder of the paper, we take $\Gamma_1$ as the energy unit.

    \begin{figure}[bt]
        \begin{center}
        \includegraphics[scale=0.48]{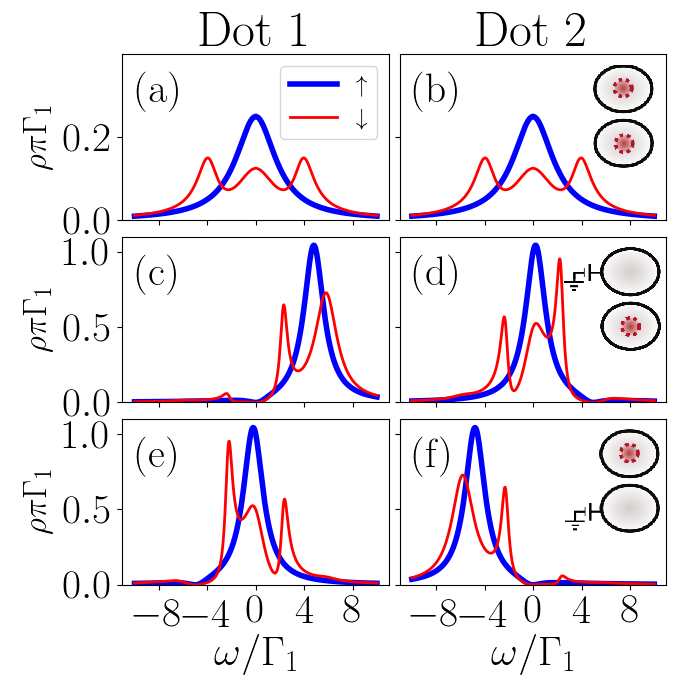}
       \caption{ \label{fig:t1EQt2}  Spin-resolved spectral densities (LDOS) $\rho_{i \sigma}(\omega)$ for noninteracting  dots $i=1,2$ in the symmetric coupling setup [Fig.\ref{fig:MajoranaModels}(a)]. Panels (a), (c) and (e) show $\rho_{1 \sigma}(\omega)$ while panels (b), (d) and  (f) depict $\rho_{2 \sigma}(\omega)$. Each row corresponds to different dot level positions $\ep_1$,$\ep_2$ controlled by gate voltages applied to each dot. (a),(b): $\ep_1=\ep_2=0$. (c),(d): $\ep_1=5\Gamma_1, \ \ep_2 =0$.  (e),(f): $\ep_1=0, \ \ep_2 =-5\Gamma_1$.  Spin-up LDOS $\rho_{i \up}(\omega)$ are marked by bold blue lines while $\rho_{i \dw}(\omega)$ are by thin red lines. Insets show where the MZM signatures, represented by a red dashed circle, are located. 
        }
        \end{center}
    \end{figure}

     \subsection{MZM manipulation in noninteracting quantum dots \label{subsect:non-int}}

\begin{figure}[bt]
\begin{center}
\includegraphics[scale=0.48]{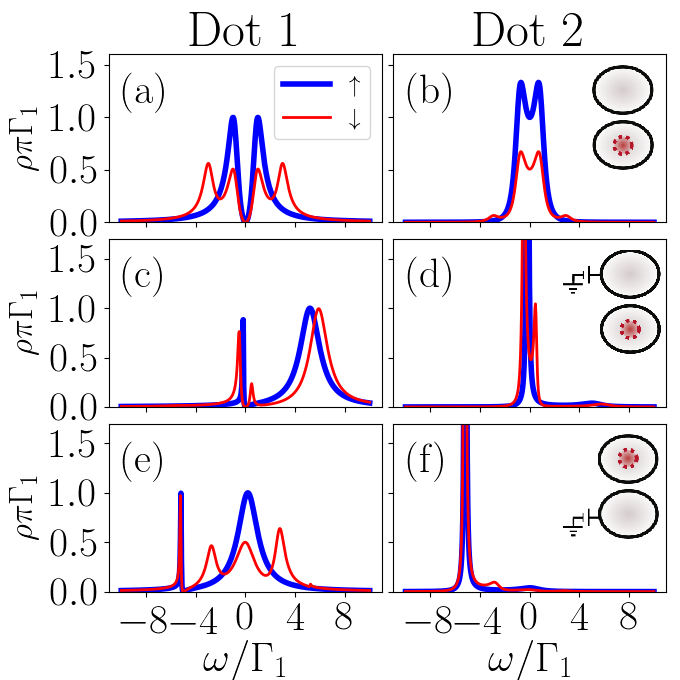}
\caption{  \label{fig:t1>0}  Spin-resolved spectral densities (LDOS) $\rho_{i \sigma}(\omega)$ for noninteracting  dots $i=1,2$ in the ``T-shaped" configuration [Fig.\ref{fig:MajoranaModels}(b)]. Panels (a), (c) and (e): $\rho_{1 \sigma}(\omega)$. Panels (b), (d) and  (f): $\rho_{2 \sigma}(\omega)$. Gate-voltage-controlled energy level  positions are identical as in Fig.\ \ref{fig:t1EQt2}: (a),(b): $\ep_1=\ep_2=0$. (c),(d): $\ep_1=5\Gamma_1, \ \ep_2 =0$.  (e),(f): $\ep_1=0, \ \ep_2 =-5\Gamma_1$.  Spin-up LDOS $\rho_{i \up}(\omega)$ are marked by bold blue lines while $\rho_{i \dw}(\omega)$ are by thin red lines. Insets show where the MZM signatures, represented by a red dashed circle, are located. 
}
\end{center}
\end{figure}

\begin{figure}[t]
\begin{center}
\includegraphics[scale=0.48]{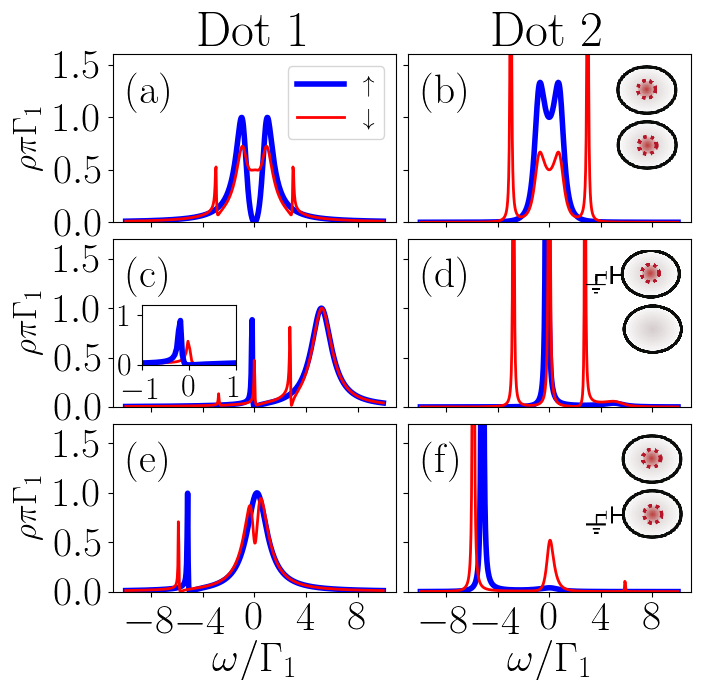}
\caption{  \label{fig:t2>0} Spin-resolved spectral densities (LDOS) $\rho_{i \sigma}(\omega)$ for noninteracting  dots $i=1,2$ in the ``in-series'' configuration [Fig.\ref{fig:MajoranaModels}(c)]. 
Panels (a), (c) and (e): $\rho_{1 \sigma}(\omega)$. Panels (b), (d) and  (f): $\rho_{2 \sigma}(\omega)$. Gate-voltage-controlled energy level  positions are identical as in Fig.\ \ref{fig:t1EQt2}: (a),(b): $\ep_1=\ep_2=0$. (c),(d): $\ep_1=5\Gamma_1, \ \ep_2 =0$.  (e),(f): $\ep_1=0, \ \ep_2 =-5\Gamma_1$.  
Spin-up LDOS $\rho_{i \up}(\omega)$ are marked by bold blue lines while $\rho_{i \dw}(\omega)$ are by thin red lines. Insets show where the MZM signatures, represented by a red dashed circle, are mainly located. Inset in (c): Magnification of the low-energy region.
}
\end{center}
\end{figure}

     The noninteracting results for setups (a),(b), and (c) of Fig.\ \ref{fig:MajoranaModels} are shown in Figs.\ \ref{fig:t1EQt2}, \ref{fig:t1>0} and \ref{fig:t2>0} respectively. In all cases, the left (right) panels depict the spectral density  of dot $1$ (dot $2$). Each row represents a different gate voltage configuration in the dots, starting with  $\epsilon_1\!=\!\epsilon_2\!=\!0$ (first row), $\epsilon_1\!=\!5\Gamma_1$, $\epsilon_2\!=\!0$ (second row), and finally $\epsilon_1\!=\!0$, $\epsilon_2\!=\!\!-5\Gamma_1$ (third row). The insets in each row show where the Majorana signature, represented by a red dashed circle inside the dot, is mainly located.

     Figure \ref{fig:t1EQt2} shows results for the symmetric coupling setup [Fig.\ \ref{fig:MajoranaModels}(a)] in the noninteracting regime. For the particle-hole symmetric case (first row), both spin-down  [$\rho_\dw(\omega)$, thin red line] and spin-up [$\rho_{i \up}(\omega)$, bold blue line] spectral densities are identical in both dots, as expected. Notice, however, that the $\rho_\dw(\omega)$ shows a three-peak structure, a consequence of the coupling with the Majorana mode. Moreover, the spin-down LDOS value at the Fermi energy is \textit{half} of the respective spin-up LDOS value $(\rho_{i \dw}(0) = \frac{1}{2}\rho_{i \up}(0))$, which signals the MZM tunneling into the dots. This Majorana signature is similar to the one observed when a single dot is coupled to a Majorana mode \cite{liu_detecting_2011,vernek_subtle_2014} and falls in our ``type-II" category mentioned above.  We thus may conclude that the MZM is delocalizing into \textit{both dots} in this symmetric configuration.

More interesting, we find that such delocalization can be reversed (and thus manipulated) by applying gate voltages in the dots. If a positive or negative gate voltage is induced in one of the dots, the spin-down LDOS at the Fermi energy can vanish at that dot while the type-I MZM signature $\rho_{i \dw}(0) = \frac{1}{2}\rho_{i \up}(0)$ remains in the other dot. This is shown in panels (c)-(f) of Fig.\ \ref{fig:t1EQt2} for the case of positive [Figs.\ \ref{fig:t1EQt2}(c) and \ref{fig:t1EQt2}(d)] and negative [Figs.\ \ref{fig:t1EQt2}(e) and \ref{fig:t1EQt2}(f)] gate voltages. 
%

    The location of the MZM signature can also be controlled by quantum interference, as illustrated in panels (a) and (b) of Fig.\ \ref{fig:t1>0}. Here, the MZM is coupled directly only to dot 1, which is then coupled to the lead, while dot 2 is coupled only to dot 1 via the interdot tunneling term, resulting in a ``side-dot'' configuration [see Fig.\ \ref{fig:MajoranaModels}(b)]. Interestingly, if the energy level of dot 2 is fixed to be in resonance with the Fermi energy of the lead, quantum interference causes the spectral function in dot 1 to \textit{vanish} at the Fermi level [Fig.\ \ref{fig:t1>0}(a)], while a type-I MZM signature appears in dot 2 only [Fig.\ \ref{fig:t1>0}(b)]. This interference-induced MZM signature in dot 2 is robust against shifts in dot 1's gate voltage, as depicted in Figs.\ \ref{fig:t1>0}(c) and (d).  While dot 1's LDOS is pinned at zero at the Fermi energy, dot 2's spin-down LDOS exhibits a robust zero-mode of height $\frac{0.5}{\pi \Gamma}$, which is a type-II MZM signature. 

This qualitative picture is radically altered when dot 2's gate voltage is shifted away from zero [Figs.\ \ref{fig:t1>0}(e) and \ref{fig:t1>0}(f)]. In this case, dot 2 is no longer in resonance with the leads, which changes the interference conditions such that dot 1 spectral function is no longer pinned at zero. The plots clearly show that the MZM signature, previously located in dot 2, now appears in dot 1. Moreover, the spin-up and spin-down LDOS in dot 1 become very similar to the spectral densities observed in the case of a single dot \cite{liu_detecting_2011,vernek_subtle_2014}, which indicates that dot 2 is essentially decoupled from the MZM.

    Finally, we consider the ``in-series'' configuration depicted in  Fig.\ \ref{fig:MajoranaModels}(c), in which is similar to the ``side-dot''  configuration [Fig.\ \ref{fig:MajoranaModels}(b)] except for the fact that the (spin-down) MZM is coupled only to dot 2. Thus, results for the spin-up LDOS are identical to those shown in Fig.\ \ref{fig:t1>0}. However, the MZM signatures in the spin-down LDOS are quite distinct. As an example, when both dots are in resonance with the lead [Figs.\ \ref{fig:t2>0}(a) and \ref{fig:t2>0}(b)], the spin-down LDOS does not vanish at $\omega\!=\!0$ as in the previous case. Instead, both dots show $(\rho_\dw(0)=\frac{0.5}{\pi \Gamma})$, which leads to MZM signatures of type-I in dot 2 and type-II in dot 1.

An important feature of the ``in-series'' geometry is that dot 1 presents a robust, gate-voltage-independent  MZM signature, despite the fact that it is not directly attached to the topological wire. As shown in Figs.\ \ref{fig:t2>0}(c) and \ref{fig:t2>0}(d), while a a shift in dot 1's gate voltage erases the MZM signature in dot $2$, it does not affect the MZM signature in dot $1$. At the same time, the MZM signature in dot 1 is robust against changes in the dot 2's gate voltage, as shown in Figs.\ \ref{fig:t2>0}(e) and \ref{fig:t2>0}(f). The only difference here is that  MZM signature types are reversed: dot 1 now shows a type-I signature while dot 2 shows a type-II one.

\subsection{Interacting dots: MZM-mediated indirect exchange \label{subsec:IndirectExchange}}

We now turn to the more realistic case of quantum dots in the Coulomb blockade regime where local electron-electron interaction terms are relevant. We consider the dots to be in an odd-$N$ Coulomb blockade valley where Kondo correlations are dominant at low-temperatures. The local Coulomb energy in the dots is accounted for by the terms $\frac{U_i}{2}(\sum_{\sigma} \hat{n}_{i\sigma}-1)^{2}$ in Eq.\ \eqref{eq:H_DQD}. For simplicity, we consider equal Coulomb repulsion energies $(U_1 \!=\! U_2 \equiv U)$ for both dots. For concreteness, the NRG calculations were performed with $U \!=\! 17.3\Gamma_1$ in both dots and a half-bandwidth of the lead electrons set at $D=2U=34.6\Gamma_1$.

Let us review some of the main features of the spectral densities of the dots in the absence of the MZM coupling. For a single dot coupled to a metallic lead, the Kondo effect is characterized by the appearance of a sharp resonance in the spectral function near the Fermi energy with a width of order $k_B T_K \!\sim\! \sqrt{U \Gamma_1} \exp \left[- \pi\frac{|\epsilon_1| |\epsilon_1 + U|}{U \Gamma_1}  \right]$. Here, $T_K \ll U$ is the Kondo temperature  of the system \cite{hewson_kondo_1997}, which will be largest at the particle-hole symmetric point (phs)  $\epsilon_1 \!=\!-\frac{U}{2}$. In the case of two dots at phs $\left(\epsilon_{di}\!=\! -\frac{U_i}{2}\right)$, both symmetrically coupled to a single lead ($\Gamma_1\!=\!\Gamma_2$), there will be an additional effective exchange interaction between the dots mediated by the lead \cite{Zitko:Phys.Rev.B:76:241305:2007,Liao:JournalofMagnetismandMagneticMaterials:377:354-361:2015}. Such exchange will compete with the antiferromagnetic Kondo coupling, producing a three-peak structure in the spectral density of both dots.

\begin{figure}[bt]
        \begin{center}
        \includegraphics[width=1.0\columnwidth]{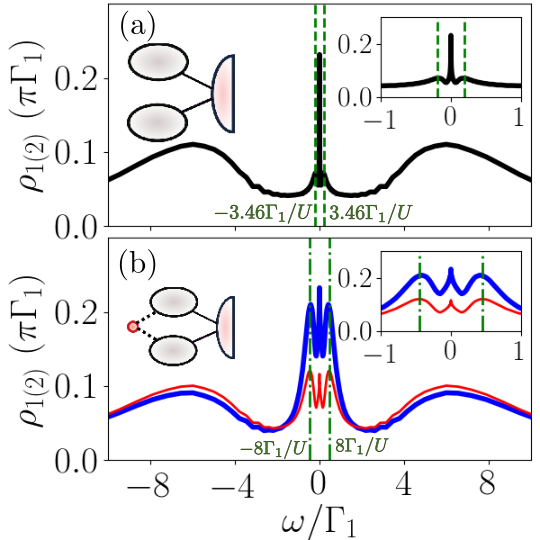}
        \caption{  \label{fig:NRG_Majorana} Spectral density (LDOS) for interacting dots ($U_1\!=\!U_2\!=\!17.3 \Gamma_1$) in the symmetric coupling configuration ($\Gamma_1\!=\!\Gamma_2$ and $t_1\!=\!t_2)$.  (a) Uncoupled MZM ($t_1\!=\!t_2\!=\!0$). Spin up and down spectral densities are identical and given by the black line. (b) Coupled MZM ($t_1\!=\!t_2\!=\!\Gamma_1$):S spin-up (bold blue lines) and  spin-down (thin red lines) spectral densities are shown. Insets: Magnification of the low-energy region. 
        }
        \end{center}
        \end{figure}

Figure \ref{fig:NRG_Majorana}(a) shows the spectral functions for both dots in this case. At large energies, the spectral density displays Hubbard peaks at $\omega \sim \epsilon_{di} \pm 8.6\Gamma_1 = \pm \frac{U}{2}$, representing the single-particle hole and electron excitations whose width is of order $\sim 4\Gamma_1$. At low energies, the spin-independent spectral densities show a central Kondo peak accompanied by indirect-exchange-induced satellite peaks at $\omega \sim \pm 3.46 \Gamma^2_1/U$, giving an energy separation that scales as $\sim \Gamma^2_1/U$ [see also insets in Fig.\ \ref{fig:NRG_Majorana}(a)].

Such exchange-driven three-peak structure remains when the MZM is coupled to the system in the symmetric coupling configuration, as shown in Figure\ \ref{fig:NRG_Majorana}(b). More striking is that the indirect-exchange splitting between the dots increases considerably with the MZM coupling up to $\sim \pm 8\Gamma_1^2/U$ : Our calculations show that the peak separation of the Majorana satellites increases quadratically with the MZM coupling $t_1\!=\!t_2$ as $4t_1^2/U$  and this effect enters in superposition with the indirect-exchange-induced satellites in Fig.\ \ref{fig:NRG_Majorana}(a). This indicates a MZM-mediated spin-spin correlation between the quantum dots. Thus, the coupling to a spin-down-polarized MZM (which is the case) affects the spin-up component of the spectral densities through this indirect spin-spin interaction. Additional details of these interesting features will be discussed elsewhere \cite{Cifuentes:inprep}.

 \subsection{MZM manipulation in interacting dots \label{subsec:Interacting}}

\begin{figure}[bt]
\begin{center}
\includegraphics[scale=0.45]{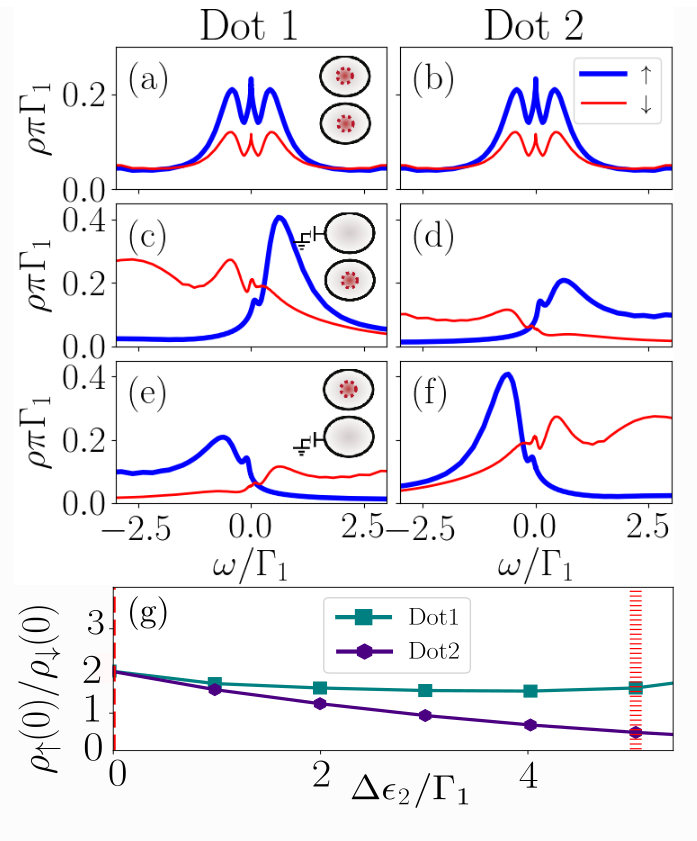}
\caption{ \label{fig:Nt1=t2} Spin-resolved spectral densities  $\rho_{i \sigma}(\omega)$ for \textit{interacting}  dots $i=1,2$ with $U_1\!=\!U_2\!=\!17.3 \Gamma_1$. Here we consider the symmetric coupling configuration shown in Fig.\ref{fig:MajoranaModels}(a). Panels (a) and (b) show $\rho_{1 \sigma}(\omega)$ and $\rho_{2 \sigma}(\omega)$ respectively for the particle-hole symmetric case $\ep_1\!=\!\ep_2\!=\!-U/2$. Panels (c) and (d) show $\rho_{1 \sigma}(\omega)$ and $\rho_{2 \sigma}(\omega)$ for $\ep_1\!=\!-U/2+\Delta \ep_1$ and $\ep_2\!=\!-U/2$ with $\Delta \ep_1=5\Gamma_1$.  Symmetrically, in panels (e) and (f), $\ep_2\!=\!-U/2+\Delta \ep_2$ and $\ep_1\!=\!-U/2$ with $\Delta \ep_2=-5\Gamma_1$ .Insets show where the MZM signatures, represented by a red dashed circle, are mainly located.  (g): Evolution of $\rho_{i \dw(0)}/\rho_{i \up(0)}$ vs $\Delta \ep_2$, for $\ep_1\!=\!-U/2$. Dashed line: $\Delta \ep_2 =0$ as in (a),(b). Barred  line: $\Delta \ep_2 =5\Gamma_1$ as in (c),(d). 
 }
\end{center}
\end{figure}

One of the key results of Fig. \ref{fig:NRG_Majorana} is that, in the symmetric configuration, both quantum dots display  type-I MZM signatures ($\rho_{i \dw}(0) = \frac{1}{2}\rho_{i \up}(0)$) co-existing with a Kondo-related zero-energy peak. A similar result was reported in Refs. \cite{lee_kondo_2013,ruiz-tijerina_interaction_2015} for the simpler case of an topological nanowire coupled to a single quantum dot. As in that case, here both Kondo and MZM signatures occur in the low-energy part of the spectral function $\omega \lesssim \Gamma_1$, as illustrated in the inset of Fig.\ \ref{fig:NRG_Majorana}(b). Within this scale, we can trace some interesting parallels with the noninteracting regime. 

As an example, Fig.\ \ref{fig:Nt1=t2} shows the NRG results for the symmetric setup in Fig.\ \ref{fig:MajoranaModels}(a). As in the noninteracting case [Fig.\ \ref{fig:t1EQt2}], type-I MZM  signatures appear in both dots.  These signatures can be manipulated by tuning the gate voltage of one of the dots to induce the MZM signature to appear only in the other dot. The LDOS at figures Fig.\ \ref{fig:Nt1=t2}(d) shows a type-I MZM  signature with $\rho_\dw(0) \approx \frac{1}{2}\rho_\up(0)$. This MZM signature is stable against gate-voltage-induced energy shifts in dot 2 away from particle-hole symmetry ($\Delta \ep_2 \equiv \ep_2+U/2$) in the range $\Delta \ep_2 \lesssim 6\Gamma_1$ (see Fig.\ \ref{fig:Nt1=t2}(e)). For larger values of $\Delta \ep_2$, dot 2 enters the mixed-valence regime and the Coulomb peak originally located at $\omega \sim \pm 8.7 \Gamma_1$ for $\Delta \ep_2\!=\!0$ now overlaps with the Fermi energy and both Majorana and Kondo signals are lost.

\begin{figure}[bt]
\begin{center}
\includegraphics[scale=0.45]{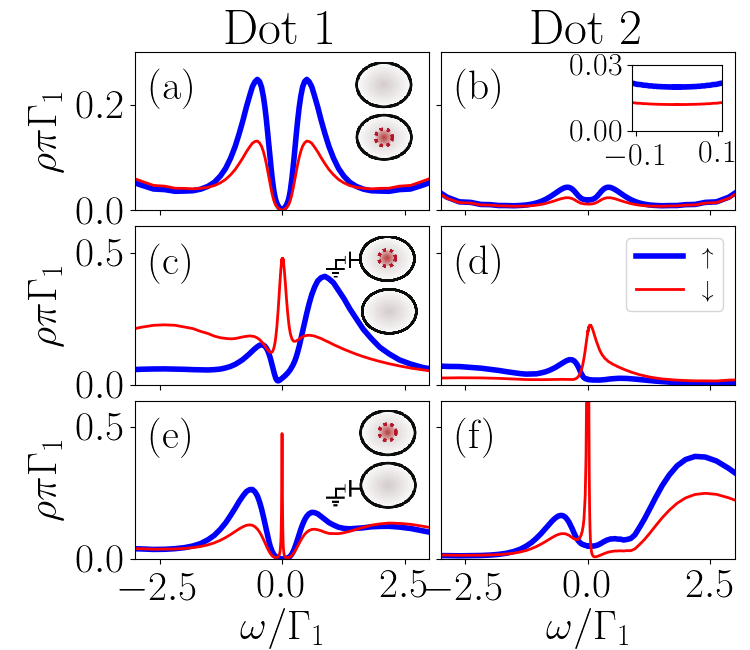}
\caption{  \label{fig:Nt1>0} Spin-resolved spectral densities $\rho_{i \sigma}(\omega)$ for interacting  dots $i=1,2$ in the ``T-shaped" configuration [Fig.\ref{fig:MajoranaModels}(b)]. Panels (a), (c) and (e): $\rho_{1 \sigma}(\omega)$. Panels (b), (d) and  (f): $\rho_{2 \sigma}(\omega)$. Energy level  positions are identical as in Fig.\ \ref{fig:Nt1=t2}: (a),(b): $\ep_1\!=\!\ep_2\!=\!-U/2$. (c),(d): $\ep_1=-U/2+5\Gamma_1, \ \ep_2 =-U/2$.  (e),(f): $\ep_1=-U/2, \ \ep_2 =-U/2-5\Gamma_1$.  Spin-up LDOS $\rho_{i \up}(\omega)$ are marked by bold blue lines while $\rho_{i \dw}(\omega)$ are by thin red lines. Insets show where the MZM signatures, represented by a red dashed circle, are mainly located. Inset in (b): Detail of the low-energy features.
}
\end{center}
\end{figure}


 Results for the interacting ``side-dot'' set-up [Fig.\ \ref{fig:MajoranaModels}(b)] are shown in Fig.\ \ref{fig:Nt1>0}. As in the noninteracting case, the spin-up spectral density of dot 1  vanishes at the Fermi level due to single-particle quantum interference, as shown in Fig.\ \ref{fig:Nt1>0}(a). In dot 2, the spectral density is drastically reduced at the Fermi level, but it remains nonzero [Fig.\ \ref{fig:Nt1>0}(b) and inset], while still showing a type-I MZM signature, namely, $\rho_{2 \dw}(0)=\frac{\rho_{2 \up}(0)}{2}$. This picture is qualitatively similar to the noninteracting case discussed previously, but it begs the question of what is the fate of the Kondo resonance in the dots in this configuration.

\begin{figure}[bt]
\begin{center}
\includegraphics[scale=0.45]{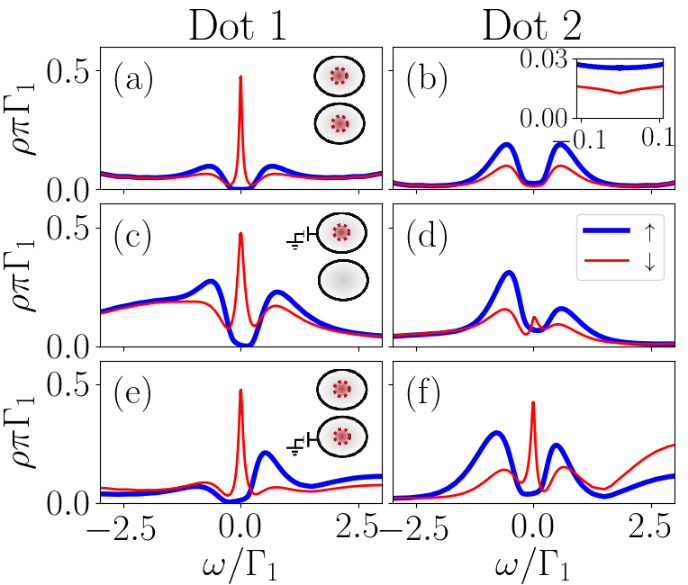}
\caption{  \label{fig:Nt2>0} Spin-resolved spectral densities $\rho_{i \sigma}(\omega)$ for interacting  dots $i=1,2$ in the ``in-series" configuration [Fig.\ref{fig:MajoranaModels}(c)]. Panels (a), (c) and (e): $\rho_{1 \sigma}(\omega)$. Panels (b), (d) and  (f): $\rho_{2 \sigma}(\omega)$. Energy level  positions are identical as in Fig.\ \ref{fig:Nt1=t2}: (a),(b): $\ep_1\!=\!\ep_2\!=\!-U/2$. (c),(d): $\ep_1=-U/2+5\Gamma_1, \ \ep_2 =-U/2$.  (e),(f): $\ep_1=-U/2, \ \ep_2 =-U/2-5\Gamma_1$.  Spin-up LDOS $\rho_{i \up}(\omega)$ are marked by bold blue lines while $\rho_{i \dw}(\omega)$ are by thin red lines. Insets show where the MZM signatures, represented by a red dashed circle, are mainly located. Inset in (b): Detail of the low-energy features.
}

\end{center}
\end{figure}

To try and answer this question, we note that a similar interplay between Kondo physics and single-particle interference on a T-shaped double dot geometry has been studied in earlier works by one of us \cite{Silva:096603:2006,dias_da_silva_transmission_2008,DiasdaSilva:Phys.Rev.Lett.:116801:2017}. It has been established that, for the case of the dot coupled to the lead (dot 1, in the present case) being noninteracting, its spectral density vanishes at the Fermi energy, while the spectral density in the second dot (dot 2) shows a ``splitted" Kondo resonance for strong enough interdot coupling. The Kondo screening in this second dot, however, is still present. In fact, the Kondo temperature \textit{increases} with the interdot coupling \cite{Silva:096603:2006,DiasdaSilva:Phys.Rev.Lett.:116801:2017}. Here the situation is slightly different as dot 1 is also interacting but we believe the analogy still holds. This picture would explain why the up and down components of the spectral density in dot 2 do not vanish at the Fermi energy (although they are quite suppressed) while still showing the MZM type-I signature ($\rho_{2 \dw}(0)=\frac{1}{2}\rho_{2 \up}(0)$).

When gate voltages are applied in either dot 1 or dot 2, a MZM signature appears in dot 1. This is shown in Figs.\ \ref{fig:Nt1>0}(c)-(f): a type-II MZM signature ($\rho_{1 \dw}(0)=\frac{0.5}{\pi \Gamma_1}$, $\rho_{1 \up}(0) \approx 0$) appears in dot 1 while neither type-I or type-II signatures are evident in dot 2. This is clearly distinct from the noninteracting case, in which a shift in the gate voltage of dot 1  [Figs.\ \ref{fig:t1>0}(c) and \ref{fig:t1>0}(d)] leads to a type-II MZM signature in dot 2 and vice-versa. When interactions are present and the system is tuned out of the particle-hole symmetric point, no clear type-I or type-II MZM signatures appear in dot 2's spectral density ([see  Figs.\ \ref{fig:Nt1>0}(d) and \ref{fig:Nt1>0}(e)]. Instead, $\rho_{2 \dw}(\omega)$ displays asymmetric resonances near the Fermi energy with comparable widths as the resonances in $\rho_{1 \dw}(\omega)$. We attribute those to Fano-like  single-particle interferences with dot 1 which are common in T-shaped structures \cite{dias_da_silva_transmission_2008}.

   Finally, Fig.\ \ref{fig:Nt2>0} depicts the NRG results for the ``series'' configuration shown in Fig.\ \ref{fig:MajoranaModels}(c). In this configuration, the MZM is coupled directly to dot 2 only. As in the noninteracting case, the most consistent MZM signatures (type-II, in this case) occur in dot 1's spectral properties.  As an illustration, Figs.\ \ref{fig:Nt2>0}(a), \ref{fig:Nt2>0}(c) and \ref{fig:Nt2>0}(e) show robust zero-energy peaks in the spin down spectral densities of dot 1 obeying $\rho_{1 \dw}(0)=\frac{0.5}{\pi \Gamma_1}$ while $\rho_{1 \up}(0) \approx 0$. The strong difference between spin up and down spectral densities clearly identifies this as a MZM signature rather than a Kondo peak. 
   
The type-II MZM signature remains in dot 2 despite changes in either $\Delta \epsilon_1$  [Fig.\ \ref{fig:Nt2>0}(c)] or $\Delta \epsilon_2$ [Fig.\ \ref{fig:Nt2>0}(e)]. Moreover,  a type-I MZM signature also appears in dot 2  in the particle-hole symmetric case, as depicted in Fig.\ \ref{fig:Nt2>0}(b). Away from particle-hole symmetry, the MZM traces in the spectral properties of dot 2 are less clear. While a shift in the dot 1 energy $\Delta \ep_1=+5 \Gamma_1$ has little effect in the dot 2 spectral density [Fig.\ \ref{fig:Nt2>0}(d)], changing the energy of dot 2 by an amount  $\Delta \ep_2=-5 \Gamma_1$ gives a zero-energy peak in $\rho_{2 \dw}(\omega)$ [Fig.\ \ref{fig:Nt2>0}(f)] which essentially meets the type-II MZM signature condition  for these parameters. 

Although this result seems to agree with the noninteracting case (see Fig.\ \ref{fig:t2>0}(f)), a closer inspection shows that this MZM signature in $\rho_{2 \dw}(\omega\!\sim\!0)$ is parameter-dependent. In fact, the height of the zero-energy peak scales roughly as $\rho_{2 \dw}(\omega\!\sim\!0) \sim (\Delta \epsilon_2)^2$ for $\Delta \epsilon_2 \lesssim 6 \Gamma_1$.  Thus, for interacting dots, the categorization of a MZM signature in dot 2 is clear only in the $4\Gamma_1 \lesssim \Delta \epsilon_2 \lesssim 6 \Gamma_1$ range, in contrast with the noninteracting case where the MZM signature is robust and largely $\epsilon_2$-independent.

One way to understand these features is to use the ``Majorana leaking'' analogy of Ref. \cite{vernek_subtle_2014}. In the series configuration of Fig.\ \ref{fig:MajoranaModels}(c), both dots can be though as nontopological ``extensions" of the Kitaev chain, with dot 1 being the ``last site'' or the ``edge''. Thus, due to the leaking of the MZM to the neighboring sites (as it is the case of a MZM attached to a single quantum dot \cite{vernek_subtle_2014,ruiz-tijerina_interaction_2015}), it would be expected that  edge-mode signatures in dot 1 would be quite robust against changes in gate voltages.


\section{Concluding remarks}
\label{sec:Conclusions}

In this paper, we have addressed the following question: Can one manipulate and detect Majorana zero modes (MZMs) in an all-electric set-up using semiconductor double quantum dots? To this end, we considered a minimal model of a MZM coupled to a double quantum dot (DQD) and metallic leads and calculated the spectral signatures in both strongly- and weakly-interacting regimes. By comparing exact analytical solutions in the noninteracting system and numerical renormalization-group results for interacting quantum dots, we were able to characterize the displacements of the MZM inside the double quantum dot for the three setups in Fig.\ \ref{fig:MajoranaModels}. 

Our results for both weakly- and strongly-interacting regime show that gate-voltage tuning in the dots allows for an effective manipulation of the tunneling of the MZM into the DQD system. By considering different MZM-DQD coupling geometries (``symmetric'', ``T-shaped'' and ``in-series'') we found that the presence or not of the MZM in each dot can be monitored by two types of signatures in the spectral density (or local density of states) of the dots.

In the symmetric configuration, the MZM is equally coupled to both dots. As in a ``double slit'' set up, the MZM signature will appear in \textit{both} dots if the gate voltages are tuned to the particle-hole symmetric (phs) point. By changing the gate voltage in one of the dots (the equivalent of ``closing one of the slits''), the MZM signature will move to the other dot. In the ``T-shaped'' configuration, when the MZM is directly coupled only to dot 1, the MZM signature will appear only in one of the dots: either dot 2 (at phs or if a gate voltage is applied to dot 1) or in dot 1, when a gate voltage is applied to dot 2. 

We also considered a configuration with the MZM coupled ``in-series'' with both dots, which is closely connected with recent design proposals for  topological quantum computational circuits involving MZMs \cite{Hoffman:Phys.Rev.B:045316:2016,karzig_scalable_2017}. In this case, there is a robust MZM signature in the ``far dot'', (the one not directly coupled to the MZM) for all gate voltage configurations, while the MZM signature in the dot directly coupled to the MZM can be manipulated via gate-voltage tuning.

Electron-electron interactions will add some interesting effects to this picture. First, there will be the appearance of a Kondo resonance in the dots, which will split due to the indirect exchange between the dots mediated by the leads.  More interestingly, we find that the coupling of the dots to the (spin-polarized) MZM will also contribute to the indirect exchange, thus creating a MZM-mediated spin exchange between the dots.  These indirect exchange effects are more prominent in the symmetric configuration, where satellite peaks in the spectral density reflect the combined Kondo-Majorana physics at low energies.

An interesting question is how robust are these features with regards to single-particle effects such as Zeeman in the QDs or the coupling of the two MZMs at the ends in the quantum wire. In order to address this issue, we have performed additional calculations to include these effects in the noninteracting limit discussed in Sec.\ \ref{sec:noninteractingMethods}. Our findings show that the overall effect of these terms is similar to the situation in the single QD coupled to a topological quantum wire \cite{liu_detecting_2011,vernek_subtle_2014,ruiz-tijerina_interaction_2015}. The energy shifts produced by Zeeman terms in the spin-resolved spectral densities are analogous to the changes in gate voltages shown in Figs.\ \ref{fig:t1EQt2},\ref{fig:t1>0} and \ref{fig:t2>0} such that the qualitative picture remains. As to the effect of the MZM-MZM coupling, we find a result similar to that of Ref. \onlinecite{liu_detecting_2011}: that spin-up and spin-down spectral densities are similar for energies lower than the coupling energy scale such that the MZM signature disappears.

As both Zeeman splitting and the coupling between MZMs are single-particle in nature, we have no reason to believe that the same qualitative picture will not hold in the interacting regime. The main effect would be a suppression of the Kondo resonance, similar to the case discussed in Ref. \onlinecite{ruiz-tijerina_interaction_2015}.

Finally, the results presented here provide a ``recipe'' for  manipulating MZMs signatures present in double quantum dots by using gate voltages and different connection geometries. This opens the interesting prospect of using multi-QD set-ups as a high-control system displaying non-Abelian braiding \cite{malciu_braiding_2018}, which can readily be integrated in recent  architecture designs for scalable topological quantum computers \cite{barkeshli_physical_2015,Hoffman:Phys.Rev.B:045316:2016,karzig_scalable_2017}.

\begin{acknowledgments}
The authors thank Edson Vernek, Ant\^onio Seridonio and Eric Andrade for enlightening discussions and suggestions.  L.~G.~G.~V.~D.~S. acknowledges financial support by CNPq (Grants Nos. 308351/2017-7 and 423137/2018-2),  and FAPESP (Grant No. 2016/18495-4). J.~D.~C. acknowledges financial support by CNPq and CAPES.
\end{acknowledgments}



 \appendix

 \section{Non-interacting Green's functions calculation \label{sec:Appendix_alg}}

 The spectral representation of the retarded Green function  \cite{zubarev_double-time_1960} associated to two fermion operators $A(t)$ , $B(t')$ is 
 \begin{equation}
 \Green{A,B} \equiv-i \int e^{i \omega t} \Theta(t)\left\langle\{A(t), B(0)\}\right\rangle d t. 
\end{equation}
Using the equations of motion technique we obtain the following relation \cite{zubarev_double-time_1960}
\begin{equation}
    \omega\Green{A,B}=\delta_{A^{\dagger},B}+\Green{\left[A,H\right],B}.
    \label{eq:Transport}
\end{equation}
\noindent We apply this expression to Hamiltonian $H$ given by Eq.\ \eqref{eq:Model}, with $B \equiv d^\dagger_{1\dw}$ and $A$ varying between the fermion operators $d^\dagger_{i\dw}, f^\dagger_\dw,c^\dagger_{k\dw}, d_{i\dw},f_\dw ,c_{k\dw}$. Taking $(\omega,t_1,t_2,\epsilon_1 \ldots)$ as fixed parameters, we obtain a closed linear system of eight  equations with eight variables of the form $\Green{A,d^\dagger_{1\dw}}$. Hence, this system has a unique solution. 

We are interested in computing an analytic expression for $\Green{d_{1\dw},d^\dagger_{1\dw}}$. The expected solution is a polynomial fraction of degree $8$, whose complexity depends on the number of couplings between the fermion operators. The method described in this paper borrows ideas from graph theory to simplify the Gauss-Jordan elimination process \cite{spielman_algorithms_2010}. We use this method to deduce a simple algorithm to solve the equations of motion of Hamiltonian $H$ of Eq.\ \eqref{eq:Model}.

Before describing the general procedure, we note that  the equations of motion of Eqs.\ \eqref{eq:Transport} for $A$ equal to $f_\dw$ and $f^\dagger_\dw$ are  
\begin{align}
        \omega\Green{f_{\downarrow},d_{1\downarrow}^{\dagger}}&= \omega\Green{f^\dagger_{\downarrow},d_{1\downarrow}^{\dagger}} \\
        &= \sum_{i=1}^2 \frac{t_i}{\sqrt{2}}\left(\Green{d_{i\downarrow},d_{1\downarrow}^{\dagger}}-\Green{d_{i\downarrow}^{\dagger},d_{1\downarrow}^{\dagger}}\right). 
\end{align}

Since $\Green{f_{\downarrow}^{\dagger},d_{1\downarrow}^{\dagger}} = \Green{f_{\downarrow},d_{1\downarrow}^{\dagger}} $ it is possible to eliminate the variable $\Green{f_{\downarrow}^{\dagger},d_{1\downarrow}^{\dagger}} $ from the system even before starting the Gauss-Jordan elimination. 

Writing the remaining EOMs in Eqs.\ \eqref{eq:Transport} for $A$ varying between  $d^\dagger_{i\dw},c^\dagger_{k\dw}, d_{i\dw},f_\dw ,c_{k\dw}$,   we obtain the following linear system
\begin{equation}
    \mathcal{T} \vec{G}_{d^\dagger_1} = \hat{e}_1,
    \label{eq:linear}
\end{equation}

\noindent where $\hat{e_1}$ is the vector with entries  $\hat{e}_{1_n} =\delta_{1n}$, $\mathcal{T}$ is the  matrix 

\begin{equation}
\left[\begin{array}{ccccccc}
\omega-\epsilon_{1} & -V_{1}^{*} & -t_{dots} & -t_{1} & 0 & 0 & 0\\
-V_{1} & \omega-\epsilon_{k} & -V_{2} & 0 & 0 & 0 & 0\\
-t_{dots}^{*} & -V_{2}^{*} & \omega-\epsilon_{2} & -t_{2} & 0 & 0 & 0\\
-t_{1}^{*} & 0 & -t_{2}^{*} & \omega & t_{2}^{*} & 0 & t_{1}^*\\
0 & 0 & 0 & t_{2} & \omega+\epsilon_{2} & V_{2}^{*} & t_{dots}^{*}\\
0 & 0 & 0 & 0 & V_{2} & \omega+\epsilon_{k} & V_{1}\\
0 & 0 & 0 & t_{1} & t_{dots} & V_{1}^{*} & \omega+\epsilon_{1}
\end{array}\right],
\label{eq:TransportMatrix}
\end{equation}

 \noindent  and $\vec{G}_{d^\dagger_1}$  is the column vector
 
 \begin{align*}
    \Big[ &  \Green{d_{\mathbf{1\downarrow}},d_{1\downarrow}^{\dagger}},\Green{c_{k\downarrow},d_{1\downarrow}^{\dagger}},\Green{d_{2\downarrow},d_{1\downarrow}^{\dagger}},\Green{f_{\downarrow},  d_{1\downarrow}^{\dagger}}, \\ & \Green{d_{2\downarrow}^{\dagger},d_{1\downarrow}^{\dagger}},\Green{c_{k\downarrow}^{\dagger},d_{1\downarrow}^{\dagger}},\Green{d_{1\downarrow}^{\dagger},d_{1\downarrow}^{\dagger}} \Big]^T.
 \end{align*}


    \begin{figure}[t]
    \begin{center}
    \centering
     \includegraphics[scale=0.29]{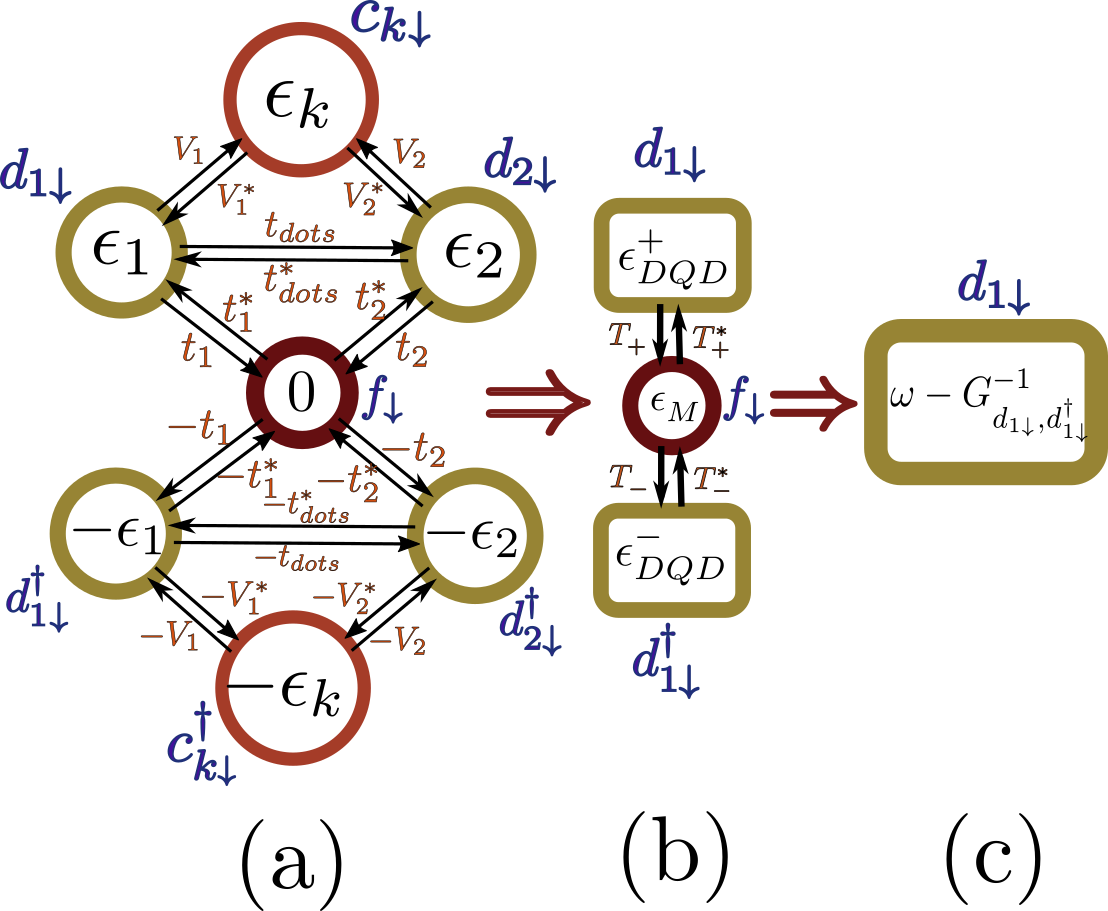}
    \caption{ Graph-Gauss-Jordan algorithm \cite{spielman_algorithms_2010}  applied to the DQD-Majorana model (a) Initial transport flow diagram  (b) Graph obtained after removing vertices $c_{k\dw},c^\dagger_{k\dw}, d_{2\dw}$ and $ d^\dagger_{2,\dw}$. New couplings in Eqs.\ \eqref{eq:epDQD+-}-\eqref{eq:M2_append} (c) Final graph after removing vertices $f_\dw, d^\dagger_{1\dw}$. The value of dot $d^\dagger_{1\dw}$ depicts the self energy of the entire system $\omega-G^{-1}_{d_{1\dw},d_{1\dw}^\dagger}$.  \label{fig:GaussJordanGraph}
    }

    \end{center}
    \end{figure}


The graph associated to the matrix given by Eq.\ \eqref{eq:TransportMatrix} is shown in Fig.\ \ref{fig:GaussJordanGraph}. Each vertex depicts the first sub-index of the Green function. The values inside each node are obtained by subtracting the corresponding diagonal term from $\omega$. We usually refer to these terms as ``self-energies". The couplings are determined by the  off-diagonal terms multiplied by $-1$.

\subsection{Solution for a DQD attached to a metallic lead}

        \begin{figure}[t]
        \begin{center}
        \includegraphics[scale=0.25]{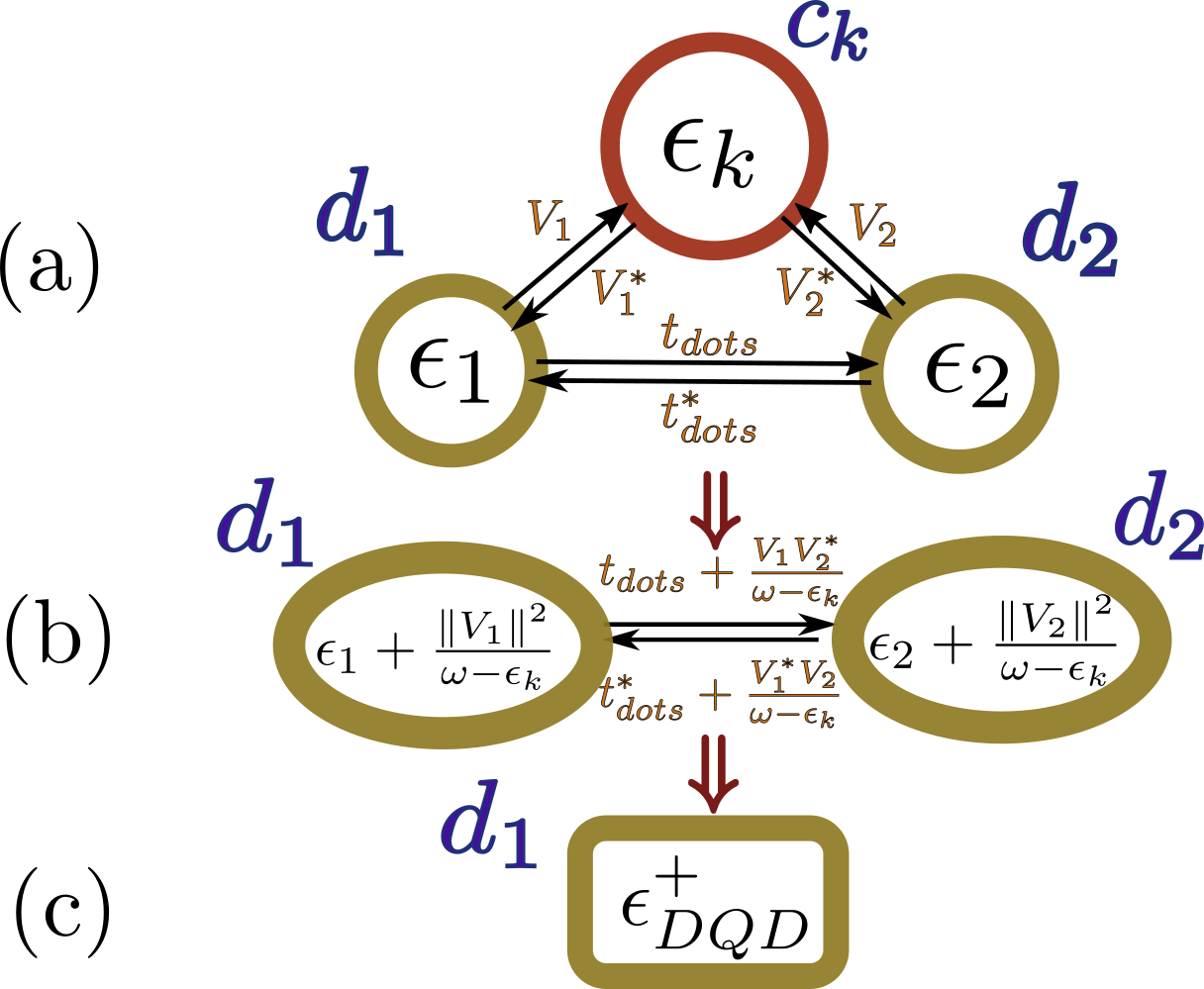}
        \caption{ Graph-Gauss-Jordan  algorithm applied  to  a DQD attached to a lead. (a) Initial transport flow diagram  (b) Graph obtained after removing vertex $c_{k\dw}$ .  (c) Remaining vertex with self energy $\ep^+_{DQD}$ .
        }
        \label{fig:GraphsDQD}
        \end{center}
        \end{figure}

Before attempting to solve the entire system, we will proceed to explain the  Graph-Gauss-Jordan \cite{spielman_algorithms_2010} elimination  process  in a  DQD-model without Majorana fermions $(t_1= t_2=0)$. This is equivalent to find the solution for the $3\times 3$ upper-left block matrix given in Eq.\ \eqref{eq:TransportMatrix} 
\begin{equation}
        \left[\begin{array}{ccc}
    \omega-\epsilon_{1} & -V_{1} & -t_{dots}\\
    -V_{1}^{*} & \omega-\epsilon_{k} & -V_{2}\\
    -t_{dots}^{*} & -V_{2}^{*} & \omega-\epsilon_{2}
    \end{array}\right], \label{eq:DQDMatrix}
\end{equation}

\noindent which can be represented by the graph in Fig.\ \ref{fig:GraphsDQD}(a). In order to eliminate the vertex $c_{k\dw}$ we just need to subtract from Eq.\ \eqref{eq:DQDMatrix} the rank-$1$ matrix that cancels the row and the column corresponding to $c_{k\dw}$. This matrix is 
\begin{equation}
        \left[\begin{array}{ccc}
    \frac{V_{1}^{*}V_{1}}{\omega-\epsilon_{k}} & -V_{1}^{*} & \frac{V_{2}V_{1}^{*}}{\omega-\epsilon_{k}}\\
    -V_{1} & \omega-\epsilon_{k} & -V_{2}\\
    \frac{V_{2}^{*}V_{1}}{\omega-\epsilon_{k}} & -V_{2}^{*} & \frac{V_{2}^{*}V_{2}}{\omega-\epsilon_{k}}
    \end{array}\right]. \label{eq:rank1}
\end{equation}
The result of Eqs.\ \eqref{eq:DQDMatrix} through  \eqref{eq:rank1} is 

\begin{equation}
        \left[\begin{array}{ccc}
    \omega-\epsilon_{1}-\frac{V_{1}^{*}V_{1}}{\omega-\epsilon_{k}} & 0 & -t_{dots}-\frac{V_{2}V_{1}^{*}}{\omega-\epsilon_{k}}\\
    0 & 0 & 0\\
    -t_{dots}^{*}-\frac{V_{2}^{*}V_{1}}{\omega-\epsilon_{k}} & 0 & \omega-\epsilon_{2}-\frac{V_{2}V_{1}^{*}}{\omega-\epsilon_{k}}
    \end{array}\right]
\end{equation}
\noindent which is mapped to the graph in Fig.\ \ref{fig:GraphsDQD}(b).

Note that it is possible to associate the correction to the energies and couplings in Fig.\ \ref{fig:GraphsDQD}(b) to the ``walks'' passing through the vertex $c_{k\dw}$.  For instance, $d_{1\dw}$'s energy $\epsilon_1$ gets an extra-term $\frac{V_{1}^{*}V_{1}}{\omega-\epsilon_{k}}$ representing an additional ``walk''  from $d_{1\dw}$ to $d_{1\dw}$ passing through  $c_{k\dw}$. The terms $V_1^*$ and $V_1$ represent a movement from $d_{1\dw}$ to $c_{k\dw}$ and vice versa, while the division by $\omega-\epsilon_{k}$ can be thought of as a penalty for  passing through $c_{k\dw}$.  The same logic applies to the  coupling terms. The correction to $t_{dots}$ is $\frac{V_{1}^{*}V_{2}}{\omega-\epsilon_{k}}$ which corresponds to a path from $d_{1\dw}$ to $d_{2\dw}$ passing through the removed vertex $c_{k\dw}$. Note that this term includes the multiplication of both couplings with the vertex divided by $\omega-\epsilon_k$. This correspondence between the energy correction and eliminated paths through the graph makes this process straightforward.

The next step is to remove the vertex $d_{2\dw}$ following the same procedure. At the end, the ``self-energy'' inside  vertex $d_{1\dw}$ will be
\begin{equation}
    \epsilon^+_{DQD}=\epsilon_{1}+\sum_{\mathbf{k}}\frac{V_{1}V_{1}^{*}}{\omega-\epsilon_{\mathbf{k}}}+\frac{\left\Vert t_{dots}+\sum_{\mathbf{k}}\frac{V_{1}V_{2}^{*}}{\omega-\epsilon_{\mathbf{k}}}\right\Vert ^{2}}{\omega-\epsilon_{2}-\sum_{\mathbf{k}}\frac{V_{2}V_{2}^{*}}{\omega-\epsilon_{\mathbf{k}}}} \label{eq:EnDQD}
\end{equation}
and the green function of $\Green{d_{1\dw}d^\dagger_{1\dw}}$ in a DQD is $\frac{1}{\omega -  \epsilon^+_{DQD}}$ [see Fig.\ \ref{fig:GraphsDQD}(c)].

\subsection{The Graph-Gauss-Jordan algorithm}

The previous method to compute the Green function  $\Green{d,d^\dagger}$ of an operator $d$ can be summarized in the following steps:

\begin{enumerate}
    \item Computing the equations of motion with the second term of the Green function fixed in the creation operator $d^\dagger$. The result is a linear system of the form $\mathcal{T} \vec{G}_{d^\dagger} = \hat{e}_1$ as in Eq.\ \eqref{eq:linear}. 
     \item  Mapping the linear system to the associated directed flow graph, labeling the vertices of the graph as $\nu_n$, with $\nu_1 = d$. The self-energy $\epsilon_{n}$ of each vertex $\nu_n$ is initialized as $\omega$ minus the corresponding diagonal term $t_{nn}$ of $\mathcal{T}$  $(\epsilon_{n} = \omega - t_{nn})$.  The coupling terms $c_{ij}$ connecting two vertices $\nu_i$ and $\nu_j$ are given by the $(i,j)$-off-diagonal terms $t_{ij}$ of the matrix $\mathcal{T}$  multiplied by $-1$ $(c_{ij} = -t_{ij})$. 
    \item Removing one-by-one the vertices of the graph, starting by the last vertex $\nu_N$.  When a  vertex $\nu_n$ is removed, an extra-term is added to each energy and coupling. These extra-terms are computed as follows:
      \begin{enumerate}
        \item Self-energy $\epsilon_{i}$: Let $c_{in}$,\ $c_{ni}$ be the coupling constants associated to the links from $\nu_i$ to $\nu_n$  and from $\nu_n$ to $\nu_i$ respectively. Note that $c_{ni} = c_{in}^*$ since the matrix $\mathcal{T}$ is hermitian. Then there is an indirect path from $\nu_i$ to itself passing through $\nu_n$. When  $\nu_n$ is eliminated , the extra-term added to $\epsilon_{i}$ is   $\frac{c_{in}c^*_{in}}{\omega-\epsilon_{n}}$. 
        \item Coupling $c_{ij}$ : Let $c_{in},\ c_{nj} \neq 0$ be the coupling constants associated to the links from $\nu_i$ to $\nu_n$  and from $\nu_n$ to $\nu_j$. Then there is an indirect path from $\nu_i$ to $\nu_j$ passing through $\nu_n$. When  $\nu_n$ is eliminated , the extra-term added to $c_{ij}$ is $\frac{c_{in}c_{nj}}{\omega-\epsilon_{\nu}}$. {}
        \end{enumerate}{}
    \ignorespacesafterend  
    This process is iterated from $n=N$ until $n=1$.
    \item The self-energy in the remaining vertex $\nu_1 = d$ is related with the green-function as $\epsilon_d = \omega -\frac{1}{ \Green{d,d^\dagger}}$.
\end{enumerate}

This algorithm is equivalent to the Gauss-Jordan elimination process used in earlier real-space decimation methods \cite{Aoki:PhysicaA::114:538-542:1982} in order to obtain noninteracting Green's functions. Our approach has two additional insights: 1) The number of operations scales linearly with the number of vertices. 2) The graph structure allows one to identify minimal and maximal cutting points which simplifies the complexity of the solution. As pointed out in previous works \cite{spielman_algorithms_2010}, selecting a good order of elimination of the vertices can improve the efficiency of the algorithm. In Fig.\ \ref{fig:GaussJordanGraph}(a), for instance, it is preferable to start eliminating the vertices at the links, $c_{k\dw}$ and $c_{k\dw}^\dagger$, each one is coupled to just two nodes. Instead, the Majorana operator $f_\dw$ will be eliminated at last since it is the one with a higher number of couplings.

\subsection{Solution for a DQD-Majorana system}
From these ideas, we can execute the graph elimination process on the model in Fig.\ \ref{fig:GaussJordanGraph}(a) .  We start by removing the vertices $c_{k\dw},c^\dagger_k, d_{2,\dw}$, and $ d^\dagger_{2,\dw}$, in that order (See Fig.\ \ref{fig:GaussJordanGraph}(b)). The energies associated to $d_{1,\dw}$ and $d^\dagger_{1,\dw}$ will be similar to those in Eq.\ \eqref{eq:EnDQD}, obtaining
\begin{equation}
    \epsilon_{DQD}^{\pm}=\pm\epsilon_{1}+\sum_{\mathbf{k}}\frac{V_{1}V_{1}^{*}}{\omega-\epsilon_{\mathbf{k}}}+\frac{\left\Vert \pm t_{dots}+\sum_{\mathbf{k}}\frac{V_{1}V_{2}^{*}}{\omega-\epsilon_{\mathbf{k}}}\right\Vert ^{2}}{\omega\mp\epsilon_{2}-\sum_{\mathbf{k}}\frac{V_{2}V_{2}^{*}}{\omega-\epsilon_{\mathbf{k}}}} \; \label{eq:epDQD+-}
\end{equation}
\noindent There is also a correction in the couplings between the Majorana mode and $d_{1,\dw}$, $d^\dagger_{1,\dw}$ given by 
\begin{equation}
    T_{\pm}=\pm t_{1}\pm t_{2}\frac{\left(\pm t_{dots}+\sum_{\mathbf{k}}\frac{V_{1}V_{2}^{*}}{\omega-\epsilon_{\mathbf{k}}}\right)}{\omega\mp\epsilon_{2}-\sum_{\mathbf{k}}\frac{V_{2}V_{2}^{*}}{\omega-\epsilon_{\mathbf{k}}}}. 
    \end{equation}

In addition there appears a self-energy $\epsilon_M$ in the  Majorana operator due to the coupling between $f_\dw$ and $d_{2\dw}$. This new term is 
\begin{equation}
    \begin{aligned}
        \epsilon_{M}=  \omega -\frac{\left\Vert t_{2}\right\Vert ^{2} } {\omega-\epsilon_{2}-\sum_{\mathbf{k}}\frac{V_{2}V_{2}^{*}}{\omega-\epsilon_{\mathbf{k}}}} 
         - \frac{\left\Vert t_{2}\right\Vert ^{2}}{\omega+\epsilon_{2}-\sum_{\mathbf{k}}\frac{V_{2}V_{2}^{*}}{\omega+\epsilon_{\mathbf{k}}}}. 
    \end{aligned}
    \label{eq:M2_append}
\end{equation}
With all the terms of the graph in Fig.\ \ref{fig:GaussJordanGraph}.(b) computed, it only remains to remove the vertices $d^\dagger_{1\dw}$ and $f_\dw$, in that order. This will lead us to the final result \eqref{eq:Green_NonInteracting}. 
\begin{equation}
    G_{{d_{1\downarrow},d_{1\downarrow}^{\dagger}}}\left(\omega\right)=\frac{1}{\omega-\epsilon_{DQD}^{+}-\frac{\left\Vert T_{+}\right\Vert ^{2}}{\omega-\epsilon_{M}-\frac{\left\Vert T_{-}\right\Vert ^{2}}{\omega - \epsilon_{DQD}^{-}}}}.
     \label{eq:2Green_NonInteracting}
\end{equation}
\noindent From this analytic expression we can compute rapidly dynamic quantities such as the density of states in the noninteracting regime. In this project, it allowed us to achieve a better understanding of the system in the different couplings, and also, to predict parameters that exhibit an interesting behavior. These parameters where simulated afterwards through NRG, which has a larger run-time. 

We introduced the Graph-Gauss-Jordan algorithm as a simple, didactic and graphical method to solve the equations of motion of quadratic Hamiltonians. We hope for its extended use in condensed matter physics.


%

\end{document}